\documentclass[12pt]{article}

\begin{document}

\pagestyle{empty}
\begin{flushright}
{IPPP/03/68\\
DCPT/03/136}
\end{flushright}
\vspace*{3mm}
\begin{center}
{\Large \bf
The Gottfried sum rule : theory vs experiment}

{\bf A.L. Kataev
\footnote{Supported in part by RFBR Grants Nos. 03-02-17047 and 03-02-17177}}
\vspace {0.1cm}

Institute for Particle Physics Phenomenology, University 
of Durham, DH1 3LE, UK and \\ 
Institute for Nuclear Research of the Academy of Sciences of Russia,
117312, Moscow, Russia
\end{center}

\begin{center}
{\bf ABSTRACT}
\end{center}
\noindent
The current status of theoretical QCD calculations
and experimental measurements of the Gottfried sum rule are discussed. 
The interesting from our point of view opened problems are summarised.
Among them is the task  of estimating the measure  of 
light-quark flavour asymmetry in possible future 
experiments. 
\vspace*{0.1cm}
\noindent
\\[3cm]
Contributed to the Proceedings of 11th Lomonosov Conference on 
Elementary Particle Physics, 21-27 August, 2003, Moscow State University,
Moscow, Russia

\newpage

\section{Introduction.}

Studies of the Gottfried sum rule of charged lepton-nucleon 
deep-inelastic scattering \cite{Gottfried:1967kk},
namely  
\begin{equation}
I_G(Q^2,0,1)=\int_0^1\frac{dx}{x}\bigg[F_2^{lp}(x,Q^2)-F_2^{ln}(x,Q^2)\bigg]
\end{equation}
can provide  the important information on the possible existence of a  
light antiquark flavour asymmetry in the nucleon sea. 
Indeed, the   NMC collaboration determination 
\cite{Arneodo:1994sh}
demonstrated that its 
experimental value 
\begin{equation}
\label{NMC}
I_{G}^{\rm NMC}(4~{\rm GeV}^2,0,1)= 0.235\pm 0.026~~~~,
\end{equation}
is significantly lower than 
the quark-parton flavour-symmetric prediction 
\begin{equation}
\label{QP}
I_G(Q^2,0,1)=\frac{1}{3}~~~.
\end{equation}
This deviation is  associated with the existence 
of a non-zero integrated light-quark flavour asymmetry defined as   
\begin{equation}
\label{FA}
FA(Q^2,0,1)=\int_0^1[\overline{d}(x,Q^2)-\overline{u}(x,Q^2)]dx~~~.
\end{equation}
In spite of 
the existence of  detailed reviews on the subject 
\cite{Kumano:1997cy, Garvey:2001yq}
we think that it is worth while to return to the consideration of the 
current status of  knowledge of different aspects
related to this sum rule.

In this report, based in part on the recent work 
of Ref. \cite{Kataev:2003en}, the 
contributions of 
QCD corrections and higher-twist effects 
to the Gottfried sum rule are  
discussed first  in the case of a flavour-symmetric sea. 
Next, the results of  its 
experimental determination
are summarised. Then we briefly 
outline  various possibilities for   determinating    the integral 
$FA(Q^2,0,1)$ from previos, present  and future data. 

\section{QCD predictions.}
Let us start by  defining the arbitrary non-singlet (NS) 
Mellin moment of the difference 
of $F_2$ structure functions of charged lepton-proton and charged 
lepton-nucleon deep-inelastic scattering (DIS):
\begin{equation}
M_n^{NS}(Q^2)=\int_0^1x^{n-2}\bigg[F_2^{lp}(x,Q^2)-F_2^{ln}(x,Q^2)\bigg]dx~~~.
\end{equation}
The moment with $n=1$,
 namely the Gottfried sum rule, can be expressed as 
\begin{eqnarray}
\nonumber
I_{G}(Q^2,0,1)&=&\int_0^1\bigg[\frac{1}{3}(u_v(x,Q^2)-d_v(x,Q^2))
+\frac{2}{3}(\overline{u}(x,Q^2)-\overline{d}(x,Q^2))\bigg]dx \\ 
&=&\frac{1}{3}-\frac{2}{3}FA(Q^2,0,1)
\end{eqnarray}
where $u_v (x,Q^2)=u(x,Q^2)-\overline{u}(x,Q^2)$ and $d_v(x,Q^2)=d(x,Q^2)-
\overline{d}(x,Q^2)$ are the valence-quark distributions and 
the measure of flavour asymmetry is related to the 
difference of  the sea-quark distributions  
$\overline{u}(x,Q^2)$ and $\overline{d}(x,Q^2)$ via Eq.(\ref{FA}). 

\subsection{Perturbative  contributions.}
Consider first  the case when the  sea is flavour-symmetric. 
In  zeroth  order of perturbation theory 
the quark-parton result of Eq.(\ref{QP}) is reproduced. However, the 
quark-gluon interactions generate  non-zero corrections to $I_G$, 
defined as
\begin{equation}
I_{G}(Q^2,0,1)=AD(\alpha_s)C(\alpha_s)~~~.
\end{equation}
The anomalous dimension term is related to the   
anomalous dimension function of the first moment 
$\gamma^{n=1}(\alpha_s)$ and to the QCD $\beta$-function, namely :
\begin{eqnarray}
\label{AD}
AD(\alpha_s)&=&exp
\bigg[-\int_\delta^{\alpha_s(Q^2)}\frac{\gamma^{n=1}(x)}{\beta(x)}dx\bigg]
=1+\frac{1}{2}\frac{\gamma_1^{n=1}}{\beta_0}\bigg(\frac{\alpha_s(Q^2)}
{4\pi}\bigg) \\ \nonumber
&&+\frac{1}{4}\bigg(\frac{1}{2}\frac{(\gamma_1^{n=1})^2}{\beta_0^2}
-\frac{\gamma_1^{n=1}\beta_1}{\beta_0^2}+\frac{\gamma_2^{n=1}}{\beta_0}\bigg)
\bigg(\frac{\alpha_s(Q^2)}{4\pi}\bigg)^2.
\end{eqnarray}
The given expansion in $\alpha_s$ 
can be obtained after taking into account 
that  $\gamma_0^{n=1}=0$ and setting  $\delta=0$.

The calculations of  $\gamma_1^{n=1}$ 
\cite{Ross:1978xk,Curci:1980uw} give the following 
result
\begin{equation}
\gamma_1^{n=1}=-4(C_F^2-C_FC_A)[13+8\zeta(3)-2\pi^2]=+2.557~~~.
\end{equation}
The coefficients of the QCD $\beta$-function are well-known:
\begin{eqnarray}
\beta_0&=&\bigg(\frac{11}{3}C_A-\frac{2}{3}f\bigg)=11-0.667f \\ 
\beta_1&=&\bigg(\frac{34}{3}C_A^2-2C_Ff-\frac{10}{3}C_Af \bigg)=102-12.667f~~~~.\end{eqnarray}
Here and below   $C_F=4/3$ and $C_A=3$ and $f$ is the number of active 
flavours.

The general order $\alpha_s^2$- expression for 
$C(\alpha_s)$ can be written down as
\begin{equation}
C(\alpha_s)=\frac{1}{3}\bigg[1+C_1^{n=1}\bigg(\frac{\alpha_s}{\pi}\bigg) 
+C_2^{n=1}\bigg(\frac{\alpha_s}{\pi}\bigg)^2\bigg]
\end{equation}
where $C_1^{n=1}=0$ \cite{Bardeen:1978yd}. The coefficient $C_2^{n=1}$ was 
evaluated   only recently  
\cite{Kataev:2003en} by means of numerical 
integration of  the  
complicated  $x$-dependence of the two-loop 
contributions to the coefficient functions of  DGLAP equation 
for  DIS structure functions, calculated  in   
Ref. \cite{vanNeerven:1991nn}.
The expression obtained in Ref. \cite{Kataev:2003en} is:
\begin{equation}
\label{C2}
C_2^{n=1}=(3.695C_F^2-1.847C_FC_A)=-0.821~~~.
\end{equation}
Collecting now all known QCD corrections to Eq.(7) we found
\cite{Kataev:2003en}:
\begin{equation}
I_G(Q^2,0,1)_{f=3}=\frac{1}{3}\bigg[1+0.0355\bigg(\frac{\alpha_s}{\pi}\bigg)
+\bigg(-0.862+\frac{\gamma_2^{n=1}}{64\beta_0}\bigg)
\bigg(\frac{\alpha_s}{\pi}\bigg)^2\bigg]~~~, 
\end{equation}
\begin{equation}
I_G(Q^2,0,1)_{f=4}=\frac{1}{3}\bigg[1+0.0384
\bigg(\frac{\alpha_s}{\pi}\bigg)
+\bigg(-0.809+\frac{\gamma_2^{n=1}}{64\beta_0}\bigg)
\bigg(\frac{\alpha_s}{\pi}\bigg)^2\bigg]~~~,
\end{equation}
where $\alpha_s=\alpha_s(Q^2)$ and the three-loop anomalous dimension term 
$\gamma_2^{n=1}$ is still unknown. In Ref.\cite{Kataev:2003en} it was 
estimated using the 
feature 
observed in Ref.\cite{Kataev:2001kk}
that  the $n$-dependence  
of the ratio $\gamma_1^{n}/\gamma_2^{n}$, obtained from 
three-loop terms  of the  
the anomalous dimension functions of even moments for charged lepton- nucleon 
DIS, calculated in Ref.\cite{Larin:1993vu}, and 
of odd moments of $\nu N$ DIS, calculated in Ref. \cite{Retey:2000nq},
can be fixed by similar approximate relation.
Taking these estimates into account we got 
\cite{Kataev:2003en}
\begin{equation}
I_G(Q^2,0,1)_{f=3}=\frac{1}{3}\bigg[1+0.0355\bigg(\frac{\alpha_s}{\pi}\bigg)
-0.811
\bigg(\frac{\alpha_s}{\pi}\bigg)^2\bigg]~~~,
\end{equation}    
\begin{equation}
I_G(Q^2,0,1)_{f=3}=\frac{1}{3}\bigg[1+0.0384\bigg(\frac{\alpha_s}{\pi}\bigg)
-0.822
\bigg(\frac{\alpha_s}{\pi}\bigg)^2\bigg]~~~,
\end{equation}
where the $\alpha_s^2$ contribution is dominated by the numerical value
of the coefficient $C_2^{n=1}$ from Eq. (\ref{C2}). Thus we convinced ourselves  that the perturbative QCD corrections  to the Gottfried sum rule 
are really small and cannot be responsible for violation of 
the  flavour-symmetric prediction  from the experimental 
value of Eq.(2).
\subsection{Higher-twist terms.}
The possibility that the higher-twist effects  
in the  Gottfried sum rule might be sizeable    was discussed 
in Ref. \cite{Szczurek:1999wp}.
It was argued that the next-to-leading sets of 
parton distributions, namely GRV94 \cite{Gluck:1994uf}, 
MRST98 \cite{Martin:1998sq} and CTEQ5 \cite{Lai:1999wy}, failed to 
describe the existing experimental  $F_2^p$-$F_2^n$ data below 
$Q^2<7~{\rm GeV}^2$ \cite{Szczurek:1999wp}.
From  the point of view of the authors 
of Ref. \cite{Szczurek:1999wp} this might be associated  
with   substantial 
higher-twist corrections, which in part are  responsible for the deviation 
of the Gottfried sum rule result from its NMC  value. However, definite 
results of the fits to $F_2^{p}-F_2^{n}$ data, performed   
in Ref.\cite{Alekhin:2003qq}
with the help of the most recent Alekhin PDF set 
of Ref. \cite{Alekhin:2002fv} (A02),  indicate that   
that  the conclusions of Ref. \cite{Szczurek:1999wp} are too optimistic.
Indeed, the authors of Ref. \cite{Alekhin:2003qq}   demonstrated 
that for the second NS moment of Eq. (5) the numerator of the $1/Q^2$ twist-4 
correction  is rather small, 
namely $H_{F_2}^{n-p}= -0.0058\pm 0.0069$ ${\rm GeV}^2$.
In view of this we  expect that the twist-4 contribution 
to the first moment, namely to $I_G$, will be  small also.
Moreover, in spite of the fact that the fits of  Ref. \cite{Alekhin:2003qq} 
reveal a definite discrepancy between   $x$-dependence found for  
$H_{F_2}^{n-p}(x)$ and the predictions of the infrared renormalon (IRR) model 
(for a review see Ref. \cite{Beneke:1998ui}), we think that 
the latter method still might  
give order-of-magnitude  estimates  of the higher-twist 
contributions. Note, that the IRR model is based in part 
on  summations of the chain of  fermion loop insertions to the
gluon propagator and is thus related to the 
large $f$-expansion of the coefficient functions.
For  the polarised Bjorken sum rule these studies 
were made in Ref.\cite{Broadhurst:1993ru} 
(for   more recent discussions see Ref.\cite{Broadhurst:2002bi}).
In the case of $I_G$ 
the flavour 
dependence does not manifest itself up to $\alpha_s^3$-corrections
(see Eq. (12) ).
Thus, we  conclude, that in comparison with polarised 
Bjorken sum rule, 
the IRR model corrections (and therefore 
higher-twist  effects) to the 
Gottfried sum rule will be damped by the additional factor $\alpha_s/\pi$.
In view of this  we think 
that 
the Gottfried sum rule cannot receive  substantial higher twist 
contributions, though the explicit demonstration of the validity of 
this statement   is still 
missing.      
\section{Experimental situation.}
The experimental determinations  of the Gottfried sum rule have 
a rather long history, summarised in the reviews of Refs.\cite{Kumano:1997cy,
Garvey:2001yq}. In fact what is really evaluated from the    
experimental data is the integral 
\begin{equation}
\label{Imin}
I(Q^2,x_{min},x_{max})=\int_{x_{min}}^{x_{max}}\frac{dx}{x}
\bigg[F_2^{lp}(x,Q^2)-F_2^{ln}(x,Q^2)\bigg]~~~.
\end{equation}
In a  more detailed analysis 
$F_2^{lp}(x,Q^2)-F_2^{ln}(x,Q^2)$ should be 
extrapolated to the 
unmeasured regions and since $F_2^{ln}(x,Q^2)$ is extracted from DIS on 
nuclei targets, 
nuclear effects should be also 
taken into account. However,  all  four experimental groups 
working on the direct determiantion  of the Gottfried sum rule 
from their experimental data,   were not able  
to achieve ideal results. Indeed, the main source  of experimental uncertainty 
results from the extrapolations of the experimental data from $x_{min}$ to 0.
Moreover, the mean $Q^2$ in the data are   dependent  on $x$ 
and it is sometimes  difficult to fix typical $Q^2$ value of 
the Gottfried sum rule (see Table 1.)
\begin{center}
\begin{tabular}{||r|c|c|c|c||}\hline
Group & $Q^2$ (GeV$^2$) & $x_{min}$ & $x_{max}$ & $I_G(Q^2,x_{min},x_{max})$ 
\\ \hline \hline
SLAC~\cite{Stein:1975yy} & 0.1--20  & 0.02 & 0.82 & 0.200$\pm$0.040 \\ \hline
EMC~\cite{Aubert:1987da}
& 10--90 & 0.02 & 0.8 & 0.197$\pm$0.011(st.)$\pm$0.083~(sys.)  \\ \hline
BCDMS~ \cite{Benvenuti:1989gs}& 20 & 0.06 & 0.8 & 0.197$\pm$0.011 (st.)$\pm$0.036(sys.) \\ \hline
NMC~\cite{Arneodo:1994sh}   & 4 & 0.004 & 0.8 & 0.221$\pm$0.008 (st.)$\pm$0.019(sys.)  
\\ \hline\hline
\end{tabular}
\end{center}
{{\bf Table 1.} The existing experimental data for the integral 
of Eq. (\ref{Imin}).}

In spite of the fact that already the results of Ref. \cite{Stein:1975yy}
inspired discussions of the possibility that the theoretical 
expression for Eq.(\ref{QP}) 
might be violated, its further determinations 
by EMC collaboration \cite{Aubert:1987da},
namely 
\begin{equation}
I_G^{\rm EMC}(Q^2=?,0,1)=0.235^{+0.110}_{-0.099}    
\end{equation}
within experimental error bars did not in fact 
demonstrate an  obvious deviation  
from the quark-parton model prediction $1/3$ 
(note, that in Eq.(19) $Q^2$-value was not determined).
A similar conclusion 
also applies  to the analysis of  
the BCDMS data in Ref. \cite{Benvenuti:1989gs}.
Indeed, it 
suffers from   large 
uncertainties at  $x<0.06$, which vary
from $0.07$ to $0.22$. Thus the results of NMC collaboration 
of  Eq.(\ref{NMC}) turned out to be extremely important for 
understanding that flavour-asymmetry of antiquark distributions in the nucleon 
really exist in nature. The precision of their data even allow one 
to extract  
the value of  integrated light-quark flavour asymmetry, defined by 
Eq.(4) \cite{Arneodo:1994sh}:
\begin{equation}
FA^{\rm NMC}(4~{\rm GeV}^2,0,1)=0.147\pm0.039~~~.
\end{equation} 
However,  even the members of NMC collaboration were 
not able to take into account all  effects, typical for DIS.  Indeed, 
the nuclear corrections, such as the Fermi motion, were neglected 
by them. In view of this, it became rather important to 
get an independent experimental extraction of $FA(Q^2,0,1)$. 
Quite recently this was done in an  analysis 
of the data for  Drell-Yan production  in proton-proton 
and proton-deuteron scattering by the members 
of E866 collaboration. Integrating   $\overline{d}(x,Q^2)-
\overline{u}(x,Q^2)$ over the measured 
$x$-region they obtained 
\cite{Towell:2001nh}:
\begin{equation}
FA^{\rm E866}(54~{\rm GeV}^2,0.015,0.35)=0.0803\pm0.011~~~.
\end{equation}
Extrapolation   this integral to the unmeasured region 
$0\leq x\leq 0.015$   
and assuming that the contribution for $x\geq 0.35$ is negligible, 
the members of E866 collaboration found that 
\begin{equation}
FA^{\rm E866}(54~{\rm GeV}^2,0,1)=0.118\pm0.015
\end{equation} 
(see Ref.\cite{Towell:2001nh}). Within existing error-bars 
this result turned out to be in agreement with Eq. (20), which is 
the  NMC value of 
this integral, extracted at 4 ${\rm GeV}^2$. In view of this it is possible 
to conclude that the value of the integrated  light-quark flavour asymmetry 
is almost independent of $Q^2$ over a wide range 
of the  momentum transfer. 
This   demonstrates in part its  non-perturbative 
origin.
\section{Possible future prospects.}
In order to calculate the characteristics of a light-quark flavour asymmetry 
a number of non-perturbative models have been successfully used. 
Among them are the meson-cloud model,  instanton model,
chiral-quark soliton models and others 
(for reviews  see Refs.
\cite{Kumano:1997cy,Garvey:2001yq}). However, to make more 
definite conclusions 
on the predictive power of these approaches more detailed experimental 
knowledge about the behaviour of the rate $\overline{d}/\overline{u}$ 
at different values of $x$ is needed.

The new 120 GeV Fermilab Main Injector should allow one  to extend 
Drell-Yan  measurements of  
$\overline{d}/\overline{u}$ to the region of $0.02<x<0.7$. 
Moreover, the studies of the CEBAF data for $F_2^D$ al large $x$ 
can give the chance to perform more detailed combined fits of all 
available DIS data.
The  extraction  
of a light-quark flavour asymmetry from the fits to the data 
for $F_2^D$ require 
the detailed treatment of nuclear effects, say in the manner of the work 
of Ref.\cite{Alekhin:2003qq}. Clearly, these measurements might be 
important for more detailed determinations 
of the effects of flavour-asymmetry 
in various  sets of parton distribution functions, which 
at present differ in the CITEQ6M, MRST2001C and A02 
sets (for their  comparison  at $Q^2=(100$ $\rm{GeV})^2$ see 
Ref. \cite{Djouadi:2003jg}).  As a test of their 
current predictive power it can be  
rather useful to use them  for calculations 
of the integral $FA(Q^2,0,1)$, as  was done in 
Ref.\cite{Garvey:2001yq}. 
 
Another  possibility is to study light-quark flavour asymmetry 
in the process of $\nu N$ DIS at the possible future Neutrino Factories.
Indeed, as was noticed in Ref. \cite{Ross:1978xk} 
in the quark-parton model the  $\overline{d}-\overline{u}$ difference can 
be related to the $\nu N$ DIS SFs as 
\begin{equation}
\label{NuFac}
\overline{d}(x)-\overline{u}(x)=\frac{1}{2}\bigg[F_1^{\overline{\nu}p}(x)-
F_1^{\nu p}(x)\bigg]-
\frac{1}{4}\bigg[F_3^{\overline{\nu}p}(x)-F_3^{\nu p}(x)\bigg]
\end{equation}
where  the $\overline{s}$ and $\overline{c}$ distributions are  neglected. 
At the NLO of perturbative QCD the analog of Eq.(\ref{NuFac}) was 
also derived \cite{Ross:1978xk}. 
Since it is known, that at the Neutrino Factories it will be possible 
to extract from the cross-sections of $\nu N$ DIS 
$F_1$ and $F_3$ structure functions 
separately \cite{Mangano:2001mj}, the more precise 
$\nu N$ DIS data might  be useful for additional 
estimates of the 
size  of the  light-quark flavour asymmetry.
  
  
\section*{Acknowledgements}
I am grateful to G. Parente for the  fruitful collaboration
and to  A. B. Kaidalov for discussions during this interesting 
Conference. It is also a pleasure 
to thank W.  J. Stirling and C. J. Maxwell  
for useful conversations  and all members of IPPP (Durham) for hospitality 
during the work on this report.

\end{document}